\documentclass[aps,prb,twocolumn]{revtex4}
\usepackage{graphicx}
\usepackage{amssymb}
\newcommand{\beq}{\begin{equation}}
\newcommand{\eeq}{\end{equation}}
\newcommand{\bea}{\begin{eqnarray}}
\newcommand{\eea}{\end{eqnarray}}
\newcommand{\beas}{\begin{eqnarray*}}
\newcommand{\eeas}{\end{eqnarray*}}
\newcommand{\nn}{\nonumber}

\newcommand{\pdag}{{\phantom{\dagger}}}

\begin{document}
\title{Soliton Fermi Sea in Models of Ising-coupled Kondo impurities}
\author{Stefan Kehrein and Matthias Vojta}
\affiliation{Theoretische Physik III, Elektronische Korrelationen und
Magnetismus, Universit\"at Augsburg, 86135 Augsburg, Germany}
\date{\today}
\begin{abstract}
We study a model of Ising-coupled Kondo impurities that can be applied to
quantum dots with capacitance coupling, coupled qubits with an incoherent
environment, etc. We show that this model becomes equivalent to a 
Anderson impurity model with a novel {\em solitonic Fermi sea}. We derive
exact results at the Toulouse point, and use the flow equation method to
extend this analysis away from the Toulouse point for not too large Ising couplings.
\end{abstract}
\maketitle
{\em Introduction.}
Kondo physics plays a fundamental role for the low-temperature behavior
of a large variety of physical systems like magnetic impurities in
metals, heavy-fermion systems, glasses, quantum dots, etc. Its key
feature is the quenching of the impurity entropy through nonperturbative
screening by many-particle excitations in the associated quantum bath.
For magnetic impurities in metals this amounts to the formation
of the Kondo singlet between the localized spin and electron--hole
excitations in the Fermi sea \cite{hewson}.

Most of the aspects of single-impurity Kondo physics are now
well understood after theoretical tools have been developed
that can deal with its intrinsic strong-coupling nature
\cite{Wilson75,Betheansatz}.
However, in many physical systems the interaction of different
impurities, i.e., multi-impurity Kondo physics, is important.
For example in heavy-fermion systems the RKKY interaction
between different impurity spins is responsible for the
competition between local Kondo physics and long-range magnetic
order that determines their phase diagram \cite{doniach}, or in glasses the
phonon-mediated coupling between different tunneling centers
is known to be important \cite{Kassner90}.
More recently, related questions about coupled two-level systems have
gained much interest in quantum computation, where decoherence due
to unwanted couplings among qubits and between qubits and environment
should be avoided; on the other hand the intentional coupling of qubits
is the key step to performing quantum logic operations.

In the present paper we investigate the case of two Kondo
impurities coupled via an $S_z$-$S_z$ interaction, i.e., an Ising-type
coupling. Equivalently, one can think of two two-level systems with
transversal coupling, with the experimental realization of coupled
flux qubits \cite{Mooij99}.
Based on the flow equation method \cite{Wegner94}
we can derive analytical results for this problem by mapping it to a
conventional Anderson impurity model. This mapping allows us to
derive exact results at the Toulouse point, and controlled approximations
in its vicinity.
Also we will show how our results can be extended
to an arbitrary number of Ising-coupled impurities with an
infinite-range interaction.

Coupled impurities or two-level systems have been investigated in a number
of papers \cite{2imp,2impnrg,2impsakai,2impcft,2impoli},
however, most attention has been focussed on the
case of SU(2)-symmetric coupling between the impurity spins.
We will show that new phenomena arise in the Ising-coupled case due to
the two-fold degeneracy of the ground state of the isolated impurity system \cite{Andrei}.
In the context of Kondo impurities, this Ising-like coupling
can be thought of as an effective impurity interaction for heavy-fermion systems
with an easy axis.
Also, Ising coupling appears naturally in quantum dots
that are coupled via their mutual capacitance:
here the two-level systems are pseudospins representing the number of electrons on the
dots, and therefore SU(2) symmetry is broken from the
outset \cite{Matveev91,Matveev95,Golden96}.

{\em Model.}
Let us first formulate the model in the language of Kondo impurities.
For simplicity, we will consider a situation where the two impurities are
placed in different fermionic baths, yielding the Hamiltonian
$H^{\rm K}=H_1^{\rm K}+H_2^{\rm K}+H_{12}^{\rm Ising}$,
where $H_i^{\rm K}$ are the individual Kondo Hamiltonians
\beq
H_i^{\rm K}=\sum_{k \alpha} \epsilon^\pdag_k\, c^\dag_{k \alpha i} c^\pdag_{k \alpha i}
+J\:\vec S^\pdag_i \sum_{\alpha \beta}
c^\dag_{0 \alpha i}\, \vec\sigma^\pdag_{\alpha\beta}\, c^\pdag_{0 \beta i}
\label{ind_Kondo}
\eeq
with the fermion creation/annihilation operators $c^\dag_{k \alpha i}$, $c^\pdag_{k \alpha i}$
with wave vector~$k=2\pi n/L$, $n\in {\mathbb Z}$, spin index~$\alpha$ and
belonging to Fermi sea~\#$i$.
$c^\dag_{0 \alpha i},~c^\pdag_{0 \alpha i}$ denote the localized electron state
at the impurity sites and $\vec S_1$, $\vec S_2$
are the two spin-1/2 impurities. We will assume a linear dispersion
$\epsilon_k=v\,k$ throughout this paper and set $\epsilon_F=0$.
The Ising coupling, which breaks SU(2) symmetry, takes the form
\beq
H_{12}^{\rm Ising}=K\,S_1^z\,S_2^z \ .
\eeq
Note that $H_{12}^{\rm Ising}$ has a doubly degenerate ground state for both signs of $K$.
The simplifying assumption of two disconnected baths is naturally satisfied for
coupled quantum dots; for magnetic impurities in metals it misses the RKKY
interaction, which we instead explicitly include by $H_{12}$.

The case of SU(2)-symmetric inter-impurity coupling, $K\,\vec S_1\cdot\vec S_2$,
has been the subject of numerous investigations in the past
\cite{2imp,2impnrg,2impsakai,2impcft,2impoli}.
Generically, two different regimes are possible as function
of the inter-impurity exchange~$K$:
for large antiferromagnetic $K$ the impurities combine to
a singlet, and the interaction with the conduction
band is weak, whereas for ferromagnetic $K$ the impurity
spins add up and are Kondo-screened by conduction
electrons in the low-temperature limit.
In both parameter regimes the ground-state spin will
be zero, and there is
{\em no} quantum phase transition as $K$ is varied
in the generic situation without particle--hole symmetry
(whereas one finds an unstable non-Fermi liquid fixed
point in the particle--hole symmetric
case)~\cite{2impnrg,2impsakai,2impcft}.
However, the finite-$T$ crossover behavior is significantly
different in the two regimes:
for large antiferromagnetic $K$, the singlet is
formed at an energy scale of order $K$.
In contrast, for ferromagnetic $K$, where both
spins have to be quenched,
a two-stage Kondo effect \cite{2imp,2impoli}
occurs if the two screening channels have
different Kondo coupling leading to different Kondo
temperatures:
upon lowering the temperature the two spin-1/2 degrees of freedom
are ``frozen'' at two distinct temperature scales.

An alternative starting point for our model of Ising-coupled impurities
can be formulated in terms
of two two-level systems (spin--boson models),
$H^{\rm SB}=H_1^{\rm SB}+H_2^{\rm SB}+H_{12}^{\rm trans}$
with
\bea
H_i^{\rm SB}&=&\frac{\Delta_i}{2}\,\sigma_i^x+\frac{1}{2}\,h^\pdag_i \sigma_i^z
+\frac{1}{2}\sum_{k>0} \lambda^\pdag_{ki}\,\sigma_i^z(b^\pdag_{ki}+b^\dag_{ki}) \nonumber \\
&&
+\sum_{k>0} \omega^\pdag_{ki}\,b^\dag_{ki} b^\pdag_{ki}
\eea
and transversal coupling between them,
$H_{12}^{\rm trans}=\frac{K}{4}\:\sigma_1^z\,\sigma_2^z$.
Here $b^\dag_{ki}$, $b^\pdag_{ki}$ are the bosonic creation/annihilation operators
for heat bath~\#$i$.
$\Delta_i$ is the bare tunneling matrix element
for the two-level system described by~$\sigma_x^i$, and $h_i$ is the asymmetry
of its two wells. The impurity properties are completely parametrized by the
spectral function
$J_i(\omega)\stackrel{\rm def}{=}\sum_k \lambda_{ki}^2\,\delta(\omega-\omega_{ki})$,
which we assume to be of Ohmic form,
$J_i(\omega)=2\alpha_i\,\omega\,e^{-\omega/2\omega_c}$. One realization
of this model is the interaction of tunneling centers in glasses through
higher-order phonon exchange \cite{Kassner90}.
In the context of quantum computation this
model arises in studies of decoherence of coupled solid state
qubits: the transversal coupling~$K$ is generated through a superconducting
flux transporter, and the heat baths describe the environment leading to
decoherence \cite{Mooij99,Wilhelm02}.

With a polaron transformation $H^{\rm SB}$
can be mapped onto the following Hamiltonian with the same impurity properties,
$H^{\rm bos}=H_1^{\rm bos}+H_2^{\rm bos}+H_{12}^{\rm trans}$.
Here
\beq
H_i^{\rm bos}=\frac{1}{2}\,h^\pdag_i \sigma_i^z
+g_i\left(V_i(\lambda_i;0)\, \sigma_i^- +{\rm h.c.}\right)
+\sum_{k>0} vk\,b^\dag_{ki} b^\pdag_{ki}
\label{bosonized_1}
\eeq
with coupling constants $g_i=\Delta_i/\sqrt{8}$
and $\omega_{ki}=vk$ without loss of generality.
$V_i(\lambda;x)$ denote vertex operators
\[
V_i(\lambda;x)\stackrel{\rm def}{=}
\exp\Big[\lambda \sum_{k>0} \frac{1}{\sqrt{k}}\,e^{-ak/2}
\big(e^{-ikx} b^\dag_{ki} - e^{ikx} b^\pdag_{ki}\big)\Big]
\]
with scaling dimension $\lambda_i=\sqrt{2\alpha_i}$
and $a=v/\omega_c$.

Eq.~(\ref{bosonized_1}) is also the bosonized form describing the low-energy
properties of the Kondo Hamiltonian~(\ref{ind_Kondo})
with $b^\dag_{ki}$, $b^\pdag_{ki}$ describing the spin-density excitations in
Fermi sea~\#$i$, $a=\rho_F v$,
$\lambda_i=\sqrt{2}-(\rho_F J)/\sqrt{8\pi^2}$, $h_i=0$, and
$g_i=J/2$.
In the sequel we will use (\ref{bosonized_1}) as the unified starting point of
our analysis and use the terminology of Kondo impurity problems. Notice
that $H_{12}^{\rm trans}$ remained unaffected by the above unitary transformation.

{\em Toulouse point.}
In a first step to study the phase diagram of Ising-coupled Kondo impurities
we investigate the special Toulouse point \cite{Toulouse69} $\lambda_i=1$
($\alpha_i=1/2$),
where we will show that the problem can be solved exactly. Since the vertex
operators $V_i(\pm 1;x)$ obey fermionic anticommutation relations,
the Hamiltonian of an individual Kondo impurity, $H_i^{\rm bos}$,
can be refermionized, leading to a resonant level
model of spinless fermions for each of the two impurities \cite{Delft98}.
The Ising coupling transforms into a density--density interaction, and
the full model $H^{\rm bos}$ thus reads
\bea
H&=&\sum_{ki} v\,k\,e^\dag_{ki} e^\pdag_{ki}
+ \sum_{ki} V_{ki}\, (e^\dag_{ki} d^\pdag_i+d^\dag_i e^\pdag_{ki}) \nn \\
&&+\epsilon_d (n_{d1}+n_{d2}-1) +\frac{h_{\rm loc}}{2}\, (n_{d1}-n_{d2}) \nn \\
&&+K\,\left(n_{d1}-\frac{1}{2}\right) \left(n_{d2}-\frac{1}{2}\right)
\label{eff_AIM}
\eea
with the representation $\sigma^+_i=d^\dag_i$, $\sigma^-_i=d^\pdag_i$,
$\sigma^z_i=2(d^\dag_i d^\pdag_i-1/2)$, $n_{di}=d^\dag_i d^\pdag_i$,
$V_{ki}=g_i\sqrt{2\pi a/L}$, $\epsilon_d=(h_1+h_2)/2$, $h_{\rm loc}=h_1-h_2$, and
\beq
e^\dag_{ki}\stackrel{\rm def}{=}\frac{1}{\sqrt{2\pi a L}}
\int dx\:e^{ikx}\: V_i(1;x) \ .
\label{solitons}
\eeq
Eq.~(\ref{eff_AIM}) represents one of the main results of our paper:
At the Toulouse point the problem of two Ising-coupled
Kondo impurities maps onto an {\em Anderson impurity model} (AIM) with
Coulomb interaction $K$. The Fermi sea of this effective AIM consists
of the {\em soliton excitations}~(\ref{solitons}) constructed from
spin-density excitations around the impurities.
The index $i=1,2$, originally labelling the two impurities and their baths,
now plays the role of a pseudospin in this effective AIM.
The impurity properties are completely described by the
hybridization function
$\Delta_i(\epsilon)\stackrel{\rm def}{=}\sum_k V_{ki}^2\,
\delta(\epsilon-\epsilon_k)$,
which is constant in the above model $\Delta_i(\epsilon)=\rho_F g_i^2$.
With this mapping the original problem (at the Toulouse point) is essentially solved
since (\ref{eff_AIM}) is well-understood using
various methods like, e.g., Wilson's numerical renormalization group~\cite{Wilson75},
and for the special case of both impurity sites being equal ($V_{ki}=V$ and
$h_{\rm loc}=0$) is even integrable with exact Bethe ansatz
results~\cite{Betheansatz}.

\begin{figure}
\centerline{\includegraphics[width=3in]{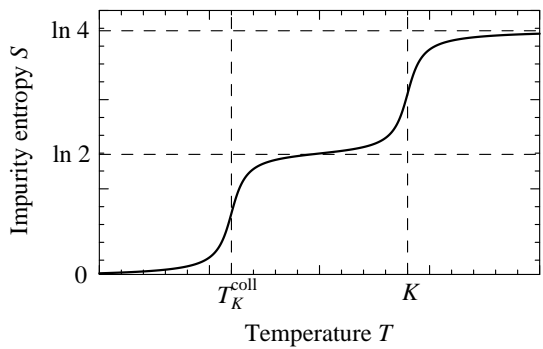}}
\caption{
Schematic temperature dependence of the total impurity entropy
of two Ising-coupled Kondo impurities for the case of
large inter-impurity coupling, $K \gg T_K^i$.
Below $K$, the impurity spins combine into the two possible
(anti)parallel low-energy states for (anti)ferromagnetic $K$.
The remaining pseudospin degree of freedom is screened below $T_K^{\rm coll}$.
}
\label{figentr}
\end{figure}

This integrable case allows us to study in detail the competition
between local Kondo screening and the Ising interaction. From the
Bethe ansatz solution one knows that the ground state of the
Anderson impurity model is always unique with no phase transition
as a function of~$K$. For simplicity
we will first analyze the problem with no local magnetic fields
$h_1=h_2=0$ ($\epsilon_d=0$).
For large positive~$K\gg \Delta(0)$
(large antiferromagnetic Ising interaction) one finds
$\langle n_{d1}+n_{d2}\rangle \rightarrow 1$
($\langle S^z_1+S^z_2\rangle \rightarrow 0$):
the impurity site of the effective Anderson model is singly
occupied.
There are two degenerate impurity configurations,
namely $\uparrow\downarrow$ and $\downarrow\uparrow$.
However, these two configurations get mixed by many-particle
excitations in the effective Fermi sea.
Thus, the pseudospin is screened below a new {\em collective Kondo
scale\/}~$T_K^{\rm coll}\sim \exp(-|K|/8\Delta(0))$,
and the ground state is a singlet. On reducing the temperature the
impurity entropy is quenched in two stages (Fig.~\ref{figentr}):
for $T\gg K$ one finds $S_{\rm imp}=\ln 4$, which is reduced to
the two singlet configurations ($S_{\rm imp}=\ln 2$) 
for temperatures of order~$K$, and finally quenched completely below
the collective Kondo temperature~$T_K^{\rm coll}$.
Note that this type of two-stage screening is completely different
from the one mentioned in the introduction \cite{2imp},
which occurs if the two conventional Kondo screening channels
have different strengths.

For large negative $K$ (large ferromagnetic
Ising interaction) the situation is related to the
antiferromagnetic case by particle--hole symmetry: one
finds $|\langle S^z_1+S^z_2\rangle | \rightarrow 1$,
i.e., the two impurity states being degenerate in the absence
of the baths are $\uparrow\uparrow$ and $\downarrow\downarrow$.
The screening of the associated pseudospin appears at the same
collective Kondo scale as above.

For smaller $|K|$, namely of order $\Delta(0)$ or below, the only relevant
temperature scale is set by $\Delta(0)$: the two impurities
do not couple into an inter-impurity doublet since they are screened
individually by their respective Kondo interaction~(\ref{ind_Kondo})
at temperatures above $K$. Notice that
at the Toulouse point $\Delta(0)$ can be identified with the Kondo
temperature $T_K^i$ of the individual impurity $i$, below which screening occurs.

Going away from the integrable case the effective Anderson impurity
model (\ref{eff_AIM}) still displays the same crossover
behavior between the various limiting cases as a function of~$K$,
provided that $K$ fulfills Eq.~(\ref{criticalK}) below.
In general the above crossover temperatures then depend on
$\epsilon_d$ and the local magnetic field~$h_{\rm loc}$.

\begin{figure}
\centerline{\includegraphics[width=3.4in]{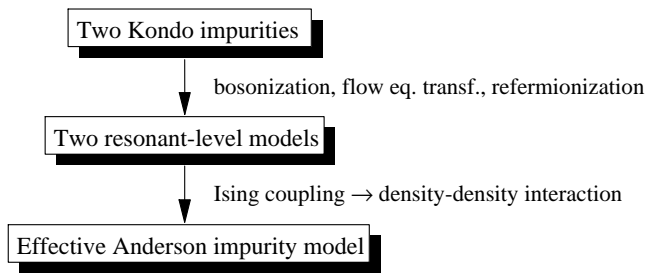}}
\caption{
Steps of our approximate solution of the two-impurity Kondo problem with
Ising coupling.
Each individual Kondo problem is transformed into an effective spinless
resonant-level model by means of a flow-equation transformation.
The original Ising coupling maps onto a density--density interaction,
leading finally to an Anderson impurity model.
}
\label{figscheme}
\end{figure}

{\em Flow equation solution.}
Next we generalize the exact results at the Toulouse point
to general couplings using the flow equation diagonalization
of the Kondo Hamiltonian~\cite{HofstetterKehrein_01}.
In Ref.~\onlinecite{HofstetterKehrein_01} a continuous sequence of
unitary transformations was constructed that led to a
systematic approximation scheme which becomes exact at the
Toulouse point and yields a controlled approximation
away from it. This means we know a unitary transformation~$U_i$
that diagonalizes the Hamiltonian~(\ref{bosonized_1}) up to
higher-order terms that do not affect the universal properties.
It was shown that
$\tilde H_i^{\rm bos}=U^\pdag_i H_i^{\rm bos}\,U^\dag_i$
takes the form
\bea
\tilde H_i^{\rm bos}&=&
\sum_{k>0} v\,k\,b^\dag_{ki} b^\pdag_{ki}
+\frac{h_i}{2}\, \tilde \sigma^z_i  \\
&&+\sum_{k} \omega_k\,[V(\lambda_i(k);k),V(-\lambda_i(k);k)]  \ , \nn
\label{diag_bosonized}
\eea
where the third term consists of Fourier transformed
vertex operators with a scaling dimension~$\lambda_i(k)$
that depends on the energy scale. The unitarily transformed
operator~$\tilde\sigma^z_i=U^\pdag_i\sigma^z_i U^\dag_i$
is given by~\cite{HofstetterKehrein_01}
\beq
\tilde\sigma^z_i=\sum_{k,k'} \alpha_k \alpha_{k'}
[V(\lambda_i(k);k),V(-\lambda_i(k');k')]
\eeq
with certain coefficients $\alpha_k$. We now make use of this
single-impurity flow equation diagonalization by applying the {\em combined}
unitary transformation $U=U_1\,U_2$ on the coupled system
\bea
\tilde H&=&U^\pdag_1 U^\pdag_2 H\, U^\dag_2 U^\dag_1 \\
&=&\tilde H_1^{\rm bos}+\tilde H_2^{\rm bos}
+\frac{K}{4}\,\tilde \sigma^z_1 \tilde \sigma^z_2 \nn \ .
\eea
This procedure is summarized in Fig.~\ref{figscheme}.
In the sequel we will only be interested in the strong-coupling regime
of the single-impurity problems $\lambda_i^2\lesssim 2$ implying
a flow to the Toulouse
point in the infrared limit $\lambda_i(k)\stackrel{k\rightarrow 0}{\longrightarrow} 1$.
We now approximate (\ref{diag_bosonized}) by setting $\lambda_i(k)=1~~\forall k$,
i.e.\ by using fermions instead of the vertex operators with a running
scaling dimension. This approximation becomes exact in the low-energy limit,
where many-particle pseudospin excitations are most important.

Using this approximation (\ref{diag_bosonized}) can be refermionized
leading to a Hamiltonian that has the same structure (\ref{eff_AIM})
like at the Toulouse point, except with a non-constant hybridization
function $\Delta_i(\epsilon)$
that depends on the unitary transformation~$U_i$. One can
show $\Delta_i(0)\sim T_K^i$, and
$\Delta_i(\epsilon)\sim \rho_F g_i^2 |\rho_F \epsilon|^{\lambda_i^2-1}$
for $|\epsilon|\gg T_K^i$, with a smooth crossover in between \cite{Slezak02}.
Here $T_K^i$ is the individual Kondo scale of one
impurity without Ising coupling to the other impurity.

The above analysis neglects the feedback of one impurity
on the other. The accuracy of the approximation within the
flow equation transformation $U_i$ decreases away from the Toulouse point.
Therefore the error in the transformation of $\tilde\sigma_i^z$ is
proportional to the distance to the Toulouse point $(\lambda-1)$,
leading to an error in the transformation of the interaction between
the two Kondo impurities that is proportional to $|K\,(\lambda-1)|$.
When this energy scale becomes comparable to the energy scale of the
individual Kondo Hamiltonians our approximations are no longer reliable,
which means that we can only study the phase diagram for
Ising couplings with
\beq
|K|\ll \frac{{\rm min}_i T^i_K}{|\lambda-1|} \ .
\label{criticalK}
\eeq
For not too large Ising couplings $K$ that fulfill this condition
we can trust the mapping to the Anderson impurity model with Coulomb
interaction $U=K$, and carry over our analysis from the Toulouse point.

For larger couplings $K$ the flow equation transformation needs to be extended
to higher orders. A careful analysis shows that a quantum phase transition
can occur when the condition (\ref{criticalK}) is violated \cite{previous_version}.
The Anderson model becomes supplemented by a density--density interaction
between the impurity orbital and the first conduction electron site.
This interaction maps onto a ferromagnetic spin--spin interaction when
charge fluctuations are frozen out for large $K$; this ferromagnetic interaction
can in turn destroy the screening of the pseudospin and lead to a
phase with a doubly degenerate ground state.
For a full account of this analysis and other approaches to the model
of Ising-coupled Kondo impurities
we refer to Ref.~\onlinecite{Garstetal}.

{\em $N$ impurities.}
The above mappings can be generalized to the case of $N$~impurities
with infinite-range Ising interaction
\beq
H=\sum_{i=1}^N H_i^K + K \sum_{i<j} S_i^z S_j^z \ .
\eeq
For simplicity we will only look at the Toulouse point
(also $h_i=0\quad \forall i$),
though the above flow equation analysis of the Kondo regime can be carried
over as well. After refermionization one finds an $N$-channel
Anderson impurity model describing the SU($N$) pseudospin
fluctuations:
\bea
H&=&\sum_{ki} v\,k\,e^\dag_{ki} e^\pdag_{ki}
+ \sum_{ki} V_{ki}\, (e^\dag_{ki} d^\pdag_i+d^\dag_i e^\pdag_{ki}) \nn \\ && \nn \\
&&+K\,\sum_{i<j} \left(n_{di}-\frac{1}{2}\right)
\left(n_{dj}-\frac{1}{2}\right) \ .
\eea

Again the crossover between ferromagnetic
and antiferromagnetic coupling~$K$ is known to be smooth.
For the special case of large positive $K$ --
this situation corresponds to a fully frustrated Ising model coupled
to dissipative baths --
one can perform a generalized Schrieffer--Wolff transformation to an
SU($N$) Kondo model that describes the quenching of the pseudospin
fluctuations below a collective Kondo scale $T_K^{\rm coll}$.
Away from the Toulouse point we find the same picture as long as
the condition (\ref{criticalK}) is fulfilled.

{\em Conclusions.}
Summing up, we have investigated the case of Ising-coupled Kondo impurities.
At the Toulouse point we could map this model {\em exactly} to an
Anderson impurity model with a Fermi sea that consists of fermionic
soliton excitations.
The Ising coupling leads to two degenerate ground states of the isolated
double-impurity system; this pseudospin degree of freedom is screened
below a collective Kondo scale $T_K^{\rm coll}$ through soliton
excitations. This piece of physics describing many-particle excitations
of solitons ({\em many-many-particle excitations})
is absent for an SU(2)-symmetric inter-impurity coupling.
Using the novel flow equation method \cite{Wegner94,HofstetterKehrein_01}
we could extend this analysis from the Toulouse point for not too
large Ising couplings $K$ that obey Eq.~(\ref{criticalK}).

This work has been supported by SFB~484 of the Deutsche
Forschungsgemeinschaft (DFG).
We acknowledge valuable discussions with
N. Andrei, M. Garst, A. Schiller, Th. Pruschke, and A. Rosch.

\end{document}